# Locality and lateral modulations of quantum well states in Ag(100) thin films studied using a scanning tunneling microscope

Takashi Uchihashi and Tomonobu Nakayama

[1]International Centre for Materials Nanoarchitectonics (MANA), National Institute for Materials Science, 1-1, Namiki, Tsukuba, Ibaraki 305-0044, Japan

**Abstract**

We investigate Ag(100) thin films epitaxially grown on a Fe(100) substrate using a low-temperature scanning tunneling microscope. Fabrication of a wedge structure by evaporating Ag through a shadow mask allows us to observe systematic evolution of quantum well (QW) states for layer thicknesses varying from 3 to 16 monolayers (ML). Close inspection of differential conductance spectra and images reveal significant modulations of QW states in the lateral directions, presumably due to the local defects at the Ag/Fe interface. The area where QW states are modulated extends over ~ 5 nm. In clear contrast, near a surface atomic step, QW states exhibit negligible changes at least up to 1 nm away from the step, leaving unmixed the two sets of neighboring QW states belonging to different thicknesses. The results are discussed in terms of a simple electron wave diffraction model.

Keywords:
quantum well states, silver, thin film, scanning tunneling microscopy, diffraction

# 1. Introduction

Quantum well (QW) states are formed when electrons are spatially confined between two parallel planes and quantized in the normal direction due to interference [1-3]. They are not only fundamentally interesting as a manifestation of quantum nature of electrons, but are also relevant to a variety of phenomena and technologies: exchange magnetic coupling in multilayers [4-6], determination of electronic band structure [7,8], stabilization of layers with magic thicknesses [9-11], modulation of superconductivity and transport properties [12-14], and quantum well/dot lasers [15], etc. State-of-the-art surface science techniques have significantly contributed to in-depth studies of QW states because they allow fabrication of samples with high-quality and well-defined surfaces and interfaces. So far, QW states have been extensively investigated for metal thin films epitaxially grown on clean substrate surfaces [4,5,7-14,16-22]. Scanning tunneling microscopy/spectroscopy (STM/S) is a powerful tool for this purpose since it can directly access their electron density of states with atomic spatial resolution. However, studies on locality and possible modulation of QW states in the lateral directions have been limited in number so far [17,18,21,22]. For example, QW states in a Pb(111) layer were resolved with a high lateral spatial resolution [17]. This was attributed to the strong anisotropy of the effective mass and resultant preferred electron motion along the surface normal direction. Nevertheless, spatial locality and possible modulation of QW states are still open questions for a metal with more isotropic electronic states.

In this paper, we study the QW states of Ag(100) thin films epitaxially grown on a Fe(100) substrate using a low-temperature (LT) STM. We observe clear evolutions of QW states for layer thicknesses varying from 3 to 16 monolayers (ML). A theoretical model based on the Bohr-Sommerfeld quantization rule successfully reproduces the experimental result. Notably, differential conductance ($dI/dV$) spectra and images reveal locally modulated QW states which extend laterally over ~5 nm. The observed bias voltage-dependent oscillatory ring patterns indicate that the phenomenon is due to the interference caused by defects at the Ag/Fe interface. In clear contrast, QW states essentially do not change up to ~1 nm away from a surface atomic step. This means that the two sets of neighboring QWS belonging to different layer thicknesses are left unmixed except within the narrow boundary regions. These seemingly contradicting results can be understood based on a simple electron-wave diffraction model.

# 2. Experimental

In the present study, Ag(100)/Fe(100) was adopted as a QW system because it possesses well-defined QW states due to an atomically sharp interface with a high reflectivity [1,7]. Compared to Ag(111) films, Ag(100) films have an advantage of having no Shockley surface states that can disturb STM observation of QW states [1]. These features make the

Ag(100)/Fe(100) system ideal for the present purpose. In the [100] direction, the Fermi surface of Ag is located near the X point in the Brillouin zone. This result in a negative effective mass at the X point in the surface normal direction while the effective mass in the lateral direction is positive [23].

Experiments were performed in an ultrahigh vacuum (UHV) chamber equipped with a LT-STM and low energy electron diffraction (LEED)/Auger electron spectroscopy (AES) apparatus. First, a single-crystal Fe(100) substrate (Matek, purity 99.98%) was sputter-cleaned by Ar ions and subsequently annealed between 300 and 750 °C [24]. This cycle was repeated for about two weeks until a clear Fe(100) LEED pattern was obtained [see Fig. 1(a)] and no AES signals except from Fe was detected. To facilitate the sputtering and annealing process, the Fe substrate was spot-welded on a thin Ta foil and was direct-current heated.

After preparing the Fe(100) surface, Ag layers was grown at ~160 K to a thickness of 5 nm by thermal evaporation. The Ag deposition was performed through a Cu mesh with a spacing of 12.5 μm (Gilder Grid, G2000HS), which was placed in front of the substrate surface [see Fig. 1(b)]. The sample was then annealed to room temperature to recover the crystallinity of Ag. This process allows us to fabricate Ag wedge structures that are readily accessible using an STM. Figure 1(c) shows a topographic STM image of a Ag wedge thus fabricated. The wedge consists of atomically flat Ag terraces that are separated by atomic steps with a height of 0.20 nm. The fact that the step height is half the Ag lattice constant (= 0.409 nm), together with formation of square-shaped islands [Inset of Fig. 1(c)], strongly indicates that Ag(100) films were epitaxially grown on the Fe(100) substrate. In Fig. 1(c), a few atomic steps originating from buried surface steps of the Fe substrate are discernible (indicated by the dashed lines) since they cross other nearly parallel Ag steps of the wedge structure.

All STM measurements were performed at 4.6 K. Sample bias voltages $V$ were measured relative to the tip. $dI/dV$ spectra were acquired by standard lock-in ac detection with a modulation amplitude $V_{\text{mod}}$ of 20 mV$_{\text{p-p}}$. $dI/dV$ images were taken with $V_{\text{mod}}$ = 10-40 mV$_{\text{p-p}}$ while scanning in the constant current mode.

## 2. Results and Discussion
## 2.1 Demonstration of quantum well states in the Ag wedge film

Figure 2(a) displays a series of $dI/dV$ spectra acquired on Ag layers with different local thicknesses ranging from 3 to 16 ML (from bottom to top). Each spectrum was obtained by averaging two to four data within the same terrace. As will be shown later, Ag surfaces were found to include mound-like local features distinguishable from other flat areas [see Fig. 3(a)]. These defects were carefully avoided to take these $dI/dV$ spectra. The spectra exhibit peaks and shoulders that systematically change their positions. They are attributed to the QW states formed within the Ag layers [1,7], since $dI/dV$ signal is proportional to $\rho(V)T(V)$ where $\rho(V)$ is the local density of states on the sample surface and $T(V)$ the tunneling transmission probability at energy $E = eV$ [25]. However, the bias voltage dependence of $T(V)$ strongly affects the $dI/dV$ spectra when taken with large voltages (here, -0.5 V < V < 2.0 V). This prevents us from precisely identifying the quantized energies. To remove this undesirable effect, the $dI/dV$ spectra were divided by $T$ to obtain normalized spectra $(dI/dV)/T$, where $T$ was obtained by fitting to each spectrum [26,27]. This process was exemplified in Fig. 2(c) where the $dI/dV$ spectrum for 16 ML was fitted with -0.5 V < V < 1.5 V. Figure 2(b) shows the results for all $(dI/dV)/T$ spectra. Now the spectra exhibit pronounced peaks instead of shoulders (indicated by triangle marks). The peak voltages $V_{peak}$ can be determined within an uncertainty of 0.05V, which result from the choice of the fitting region. The four series of $V_{peak}$ systematically increase with increasing local layer thickness $d$ [Fig. 2(d)]

This behavior is readily attributed to the energy dispersion in the [100] direction near the X point of the Brillouin zone [1,2,6,20,28]. The effective mass $m_{eff}$ near the X point in this direction is negative and the electron wave function is spatially modulated by an envelope function with a wave number $k_{env} = k_{BZ} - k$ ($k$: wave number of the electron, $k_{BZ}$ : wave number at the Brillouin zone). QW states are formed when $k_{env}$ satisfies the following Bohr-Sommerfeld quantization rule:

$$d = \left[n - 1 + \frac{\Phi(E)}{2\pi}\right]\frac{k_{BZ}}{k_{env}} \qquad (1)$$

where $\Phi(E)$ is the total phase shift caused by the reflections at the surface and the interface, and $n$ is an integer. In the two-band model, the wave number $k_{env}$ is given by [29]

$$k_{env} = \left(\frac{2|m_{eff}|}{\hbar^2}\right)^{1/2}\left\{\frac{\hbar^2 k_{BZ}^2}{|m_{eff}|} - U + E - E_{vbm}\right.$$

$$\left. - \left[U^2 + 4\left(\frac{\hbar^2 k_{BZ}^2}{2|m_{eff}|} - U + E - E_{vbm}\right)\frac{\hbar^2 k_{BZ}^2}{2|m_{eff}|}\right]^{1/2}\right\}^{1/2} \qquad (2)$$

where $U$ and $E_{vbm}$ are half the energy gap and the maximum energy at the zone boundary (X point), respectively. Here, $E$ and $E_{vbm}$ are measured relative to $E_F$. We calculated the Ag layer thickness $d$ that satisfies the $n$-th quantization condition as a function of $E$ (= $eV$). Parameters $U$ = 3.033 eV, $E_{vbm}$ = 1.721 eV, and $|m_{eff}|$ = 0.759 $m_e$ were adopted from the literature [1] and a simple linear relation $\Phi(E) = \pi$ (1.0 eV$^{-1}\times$ $E$-0.8) was assumed to obtain the best fitting. The result reproduced the experimental data excellently except for the series of $n$ =1 [Fig. 2(d), solid curves]. The deviation for $n$=1 may be derived from a likely slight non-linearity of the phase-energy relation $\Phi(E)$. Thus we have demonstrated the presence of QW states in our Ag wedge sample, which are well described within the framework of the existing theory. In the following, we will use spectra normalized by the transmission probability, (*dI/dV*)/*T*, for data presentation and discussion.

### 2.2 Lateral modulation of quantum well states near local defects

As mentioned in the previous subsection, mound-like features were present on the surface of the Ag films. Figure 3(a) shows a topographic STM image of a Ag layer with a thickness of 11 ML with such regions. They are unlikely to be purely structural defects because their topographic appearances were bias-voltage dependent. This dependence was clearly visible in *dI/dV* images [Fig. 3(b)-(e)]. At *V* = 2.0 V, the mound-like features appeared bright in the image (corresponding to high *dI/dV* signals), but they changed their contrasts as *V* was changed from 2.0 to 0.25 V (for example, see the features marked by the dashed circles). Notably, ring patterns emerged for some voltages at the rims of the mounds [Fig. 3(c)(d)]. Local *dI/dV* spectra taken across the marked region [Fig. 3(a)] also showed a clear site dependence in accordance with the above findings [Fig. 3(g)]. At the locations far away from the mound (*A, B, G, H*), the spectra exhibited peaks at *V* = 0.25, 1.1, 1.5 V, as previously found in Figs. 2(b)(d). At the locations near the rim (*C, F*), the peak at *V* = 1.1 V was strongly suppressed. Within the mound region (*D, E*), this peak was recovered but shifted by about -0.1 V. Simultaneously, the peak at *V* = 1.5 V was shifted by +0.2 V and the peak at *V* = 0.25 V almost disappeared. These findings are consistent with the changes in the *dI/dV* images in Figs. 3(b)-(e). The region for the local QW state modulation was found to extend over ~5nm.

The above observations clearly indicate the involvement of the electron interference effect [30,31]. Here, QW states already formed in the Ag layer are locally modulated due to the presence of defects below the surface. In this case, the defects were likely to be created at the Ag/Fe interface. As seen in Fig. 3(f), STM images of as-prepared Fe(100) surfaces have small dark spots, which have been attributed to oxygen impurities [24]. They should adversely affect the epitaxial growth of Ag layers because the excellent matching between Ag(100) and Fe(100) lattices (0.8%) is locally lost. The identity of the defects is not known at the present, but it may be voids formed locally on top of the Fe surface. This is inferred from the observation that the spacing between the QW state levels were increased within the mound region; in Fig. 3(g), the peaks at *V* = 1.1, 1.5 V observed outside of the mound (*A, B, G, H*) were shifted towards *V* ~ 1.0, 1.6 V within the mound (*D, E*). This suggests that the Ag layer thickness is locally reduced, consistent with formation of a void structure. Ag lattice deformations induced by a void may also contribute to the change in topographical appearance observed in Fig. 3(a).

### 2.3 Locality of quantum well states near an atomic step

In the previous subsection, we have observed that QW states are modulated in the lateral directions due to the local defects at the Ag/Fe interfere. Another location where QW states could be strongly affected is the neighborhood of a surface atomic step. This is because the Ag film thickness changes by one atomic-layer height and, consequently, the quantized energies are shifted. A naive expectation would be that the QW states are then considerably mixed and smeared out near the boundary due to the lateral electron motions. However, the degree of the "sharpness of boundary" is an open question.

To investigate this problem, we performed *dI/dV* spectroscopy near atomic steps of the Ag wedge layers. Figure 4(a) shows a series of *(dI/dV)/T* spectra taken across layers with 6 and 7 ML. The spectral locations are shown in the STM image in the upper panel. Peaks corresponding to QW states on either side of the step were clearly visible up to 0.75 nm away from the center of the step (*A, B, D, E*). The peaks were suppressed only at the middle of the step profile (*C*). This observation was corroborated by a *dI/dV* image taken across layers with 7 and 8 ML at $V$ = -0.2 V [Fig. 4(c)]. Here the *dI/dV* signal remained constant up to the very edge of the terrace except for local variations, which are ascribed to the subsurface defects as explained above. Note that the bright and dark contrasts along the step are artifacts due to a finite scanning speed for the *dI/dV* imaging. Even for thicker layers of 15 and 16 ML, QW states persisted near the step. The *(dI/dV)/T* spectra in Fig. 4(b) show that the QW state peaks were clearly visible at locations 1.1 nm away from the center of the step (*A, E*) (for the spectral sites, see the upper panel). At locations 0.55 nm away (*B, D*), the peaks were still present but slightly suppressed. At the middle of the step profile (*C*), the peaks nearly vanished. To summarize, for layers with thicknesses $d \leq 16$ ML, well-defined QW states persist up to at least 1 nm away from the step. This finding is against the naive expectation that QW states should be mixed near the step. It is also seemingly contradictory to the above observation that QW states are modulated over ~5 nm in the lateral direction. Since a 16-ML-thick Ag layer corresponds to a thickness of 3.27 nm, the lateral locality scale of 1 nm is sufficiently smaller than the vertical dimension of the layer.

## 2.3. Discussion

Here we discuss our experimental results in terms of a simple electron-wave diffraction model. Figure 5(a) shows a schematic illustration of scattering of electron waves in a QW layer. Without the defect, parallel waves propagate in the normal direction and form standing waves when the constructive interference condition is satisfied. If one introduces a point-like defect at point *O* at the bottom, a part of the incoming electron waves is scattered into spherical waves. Generally, this scattered wave has a different phase compared to that of the rest of the wave. This allows constructive interference at a certain surface point *Q* with a wave propagating from a bottom point *P*. Thus, QW states are modulated to form interference ring patterns. This explains the modulation of QW states by a local defect at the Ag/Fe interface, as observed in Fig. 3(b)-(e). The picture described here is reminiscent of diffraction of a plane wave due to a narrow hole within a barrier, where the QW states play the role of the incoming plane wave and the local defect the hole. The diffraction of waves is significant when the size of the scatterer is smaller than the wavelength. The wavelength $\lambda$ in a QW layer is roughly estimated to be $\lambda = 2d/n$, where *d* is the layer thickness and *n* is an integer quantum number. For the present system, the parameters $d = 1.2\sim3.2$ nm (6~16 ML) and $n = 1\sim4$ lead to $\lambda = 0.6 \sim 6.4$ nm. The lateral size of the local defects is estimated to be 1 nm from the STM image of the Fe(100) surface in Fig. 3(f). Hence, the requirement that the scatterer size be smaller than the wavelength can be easily fulfilled for a large *d* or a small *n*.

In contrast, electron wave scattering in QW layers with different thicknesses is schematically shown in Fig. 5(b). For simplicity, only the left layer is assumed to have well-defined QW states due to resonance, while the right layer lacks QW states due to off-resonance. The parallel incoming electron waves in the left layer are scattered at a number of points on the bottom into spherical waves, which partly propagate into the right layer. However, since the scattering occurs within a large area, the waves directed into the neighboring layer destructively interfere and becomes weak. The situation is analogous to a plane wave incident on a wide hole; as far as the hole size *L* is much larger than the wavelength $\lambda$, the diffraction is strongly suppressed.

If one supposes that diffraction is detected at a sufficiently far site, the amplitude *A* of electron waves propagating at a small angle $\theta$ measured from the normal direction is expressed as

$$A(\theta) \propto \int_0^L e^{i\Delta\phi(x)} dx = \int_0^L e^{i2\pi\theta x/\lambda} dx \qquad (3)$$

where $\Delta\phi(x) = 2\pi\theta x/\lambda$ is the relative phase of a wave scattered from point *P'*, which is at the distance of *x* from the boundary point *O'* [for schematic illustration, see the solid arrows in Fig.5(b)]. The condition $A(\theta) = 0$ is first fulfilled for $\theta L/\lambda =1$. Thus the angle $\theta$ for suppressing the diffraction is given by $\theta = \lambda/L$. If one takes the typical terrace size of the Ag wedge, *L* is estimated to be 20 nm [see Fig. 1(c)]. Then one obtains $\theta = 0.32$ for the longest wavelength $\lambda = 6.4$ nm, which corresponds to $d = 3.2$ nm and $n = 1$. Thus, QW states formed on one side of the atomic step can influence the states on the other side only up to the length of $d\theta \sim 1$ nm. For a smaller *d* and a larger *n*, the value is less than 1 nm. This is approximately equal to the lateral locality scale of 1 nm found in our experiment. For the actual sample, Eq.(3) is not accurate because a probing point *Q'* on the surface is close to the diffraction points. In this case, a lateral shift in scattering point *P'* results in an even larger phase shift due to the change in path length from *O'Q'* to *P'Q'* [see the broken arrows in Fig.5(b)]. Therefore, the destructive interference effect described above is actually more effective.

## 3. Conclusion

We have experimentally studied the locality and lateral modulation of the Ag(100)/Fe(100) QW system by taking *dI/dV* spectra and images using a LT-STM. We found significant modulations of QW states in the lateral directions, which are attributed to the interference caused by subsurface local defects at the Ag/Fe interface. The area with modulated QW states extends over ~ 5 nm. In contrast, QW states remain unchanged up to ~1 nm away the step, leaving unmixed the two sets of neighboring QW states. The results were discussed based on a simple diffraction model of electron waves. Quantitative description of QW state formation in an inhomogeneous system treated here is beyond the scope of the present research, and will be the subject of a future study.

This work was financially supported by JSPS under KAKENHI Grants No. 25247053 / 25286055/25400385 and by World Premier International Research Center (WPI) Initiative on Materials Nanoarchitectonics, MEXT, Japan. T. U. thanks for R. Berndt for fruitful discussions.

**Figure Caption**

Fig. 1 Fabrication and characterization of Ag(100)/Fe(100) samples. (a) LEED image of the Fe(100) substrate cleaned by repeated sputtering and annealing. (b) Schematic illustration of the sample used in the experiment. Ag(100) films with wedge structures were epitaxially grown on a Fe(100) substrate. (c) Topographic STM image of a Ag (100)/Fe(100) wedge (Size: 334 nm × 250 nm, Sample voltage: $V$ = 1.0V). The numbers labeled on terraces indicate the thickness of Ag layer in unit of monolayer (ML). The broken lines indicate the locations where the atomic steps of the Fe substrate are buried. The upper-right inset shows the derivative STM image of Ag layers with $d$ = 21, 21 ML (Size: 30 nm × 30 nm).

Fig. 2 QW states observed by local spectroscopy measurements. (a) $dI/dV$ spectra measured on Ag(100) layers on Fe(100) with different thickness $d$, which ranges from 3 (bottom) to 16 ML (top). Two to four spectra on the same terrace are averaged to obtain the data. (b) $dI/dV$ spectra normalized by the transmission probability $T$. The data are identical to those in (a). (c) $dI/dV$ spectrum for $d$=16 ML chosen from (a) (dotted line) and the transmission probability $T$ obtained from the fitting analysis (solid line). Fitting region: -0.5 V < $V$ < 1.5 V. (d) Peak voltages $V_{peak}$ plotted against layer thickness $d$. $V_{peak}$ were determined from the peak positions in (b). The solid lines are fits to a theoretical analysis based on the envelop functions near the Fermi level (see the text).

Fig. 3 Lateral modulations of QW states caused by local defects at the interface. (a) Topographic STM image of Ag(100) layers ($V$ = 2.0 V). The central terrace has a thickness $d$ = 11ML. The dots indicate the locations where $dI/dV$ measurements were taken [see (g)]. (b)-(e) $dI/dV$ images of the same sample region as in (a). (b) $V$= 2.0 V (c) $V$= 1.5 V (d) $V$= 1.1 V (e) $V$= 0.25 V. The dashed circles in (a)-(e) corresponds to the location of the ring shape in (c). (f) Typical topographic STM image of a Fe(100) surface ($V$ = 0.5V). (g) $dI/dV$ spectra measured at the locations *A-H* marked in (a).

Fig. 4 Locality of QW states near an atomics step. (a) Topographic STM image of Ag(100) layers with thicknesses of 6 and 7 ML (top) and $dI/dV$ spectra taken on the marked locations *A-E* (bottom). (b) Topographic STM image of Ag(100) layers with thicknesses of 15 and 16 ML (top) and $dI/dV$ spectra taken on the marked locations *A-E* (bottom). (c) $dI/dV$ images of Ag(100) layers with thicknesses of 7 and 8 ML ($V$ = -0.2 V).

Fig. 5 Schematic illustration of electron-wave diffraction in QW layers. (a) Uniform QW layer

with a local defect at the bottom. Electron waves are scattered by a local defect and the QW states are locally modulated. (b) Two neighboring QW layers with different thicknesses. Electron waves scattered at the bottom of the left QW layer interfere with themselves destructively, suppressing diffraction towards the right QW layer.

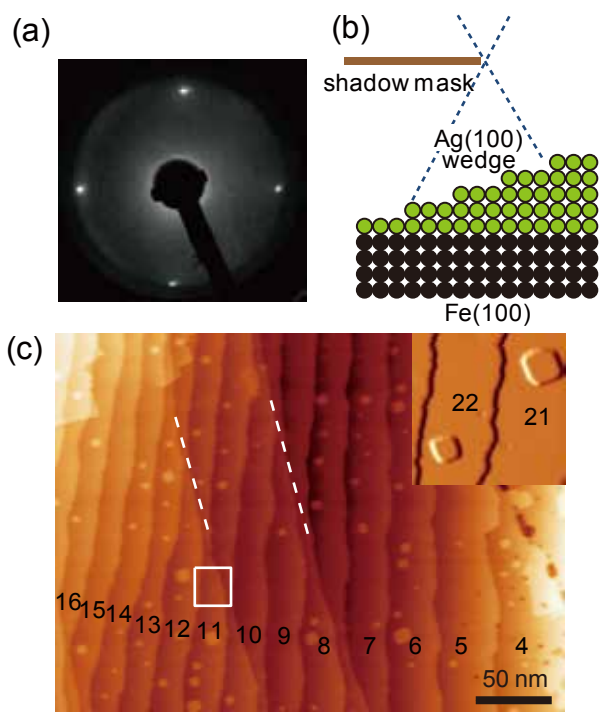

Fig. 1

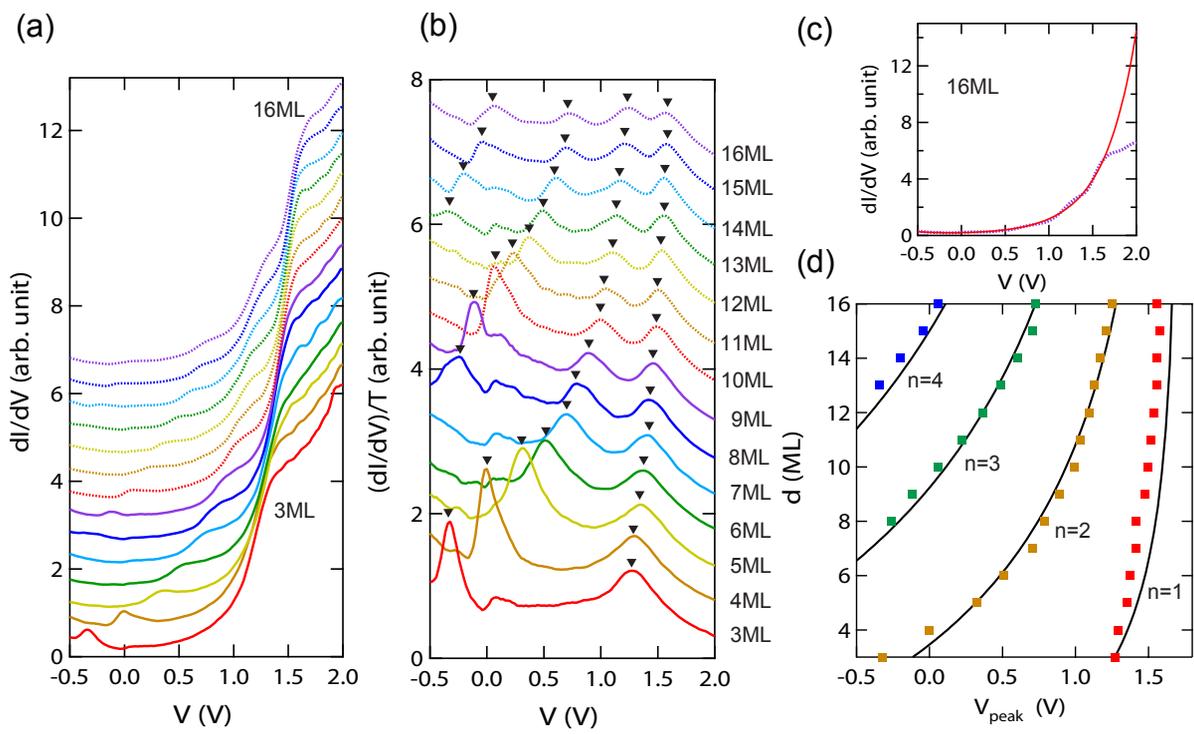

Fig. 2

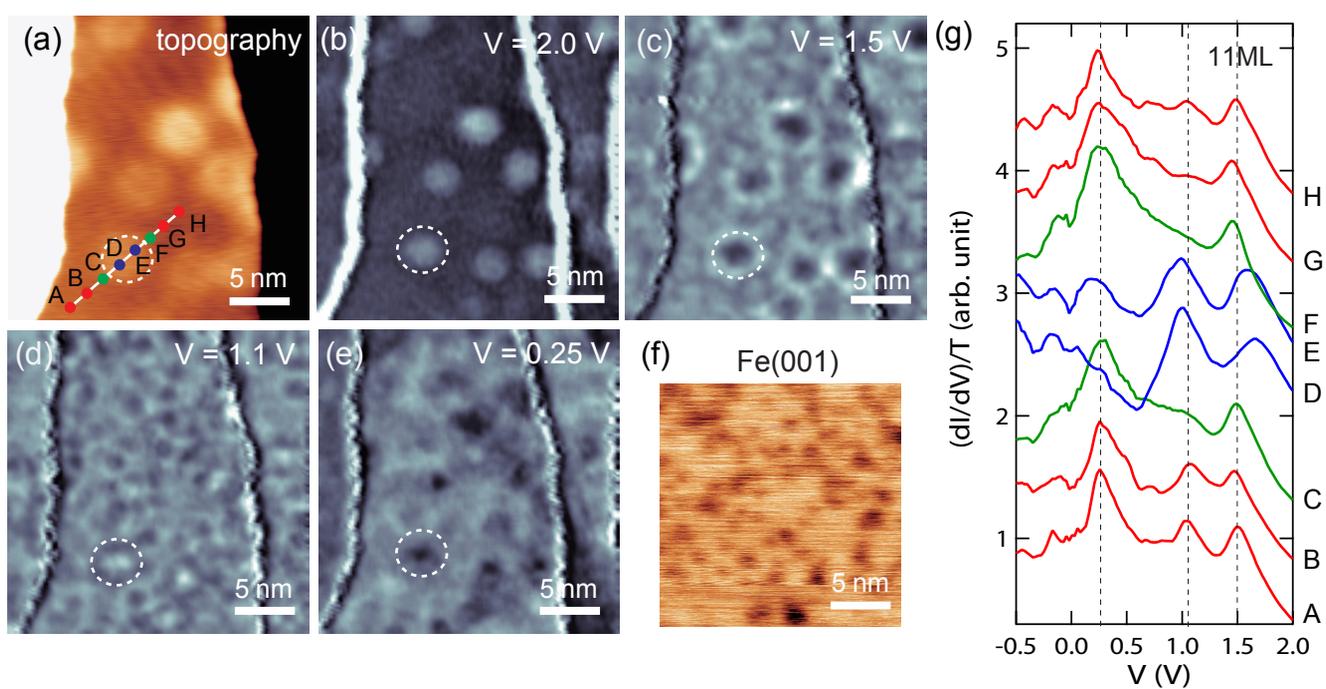

Fig. 3

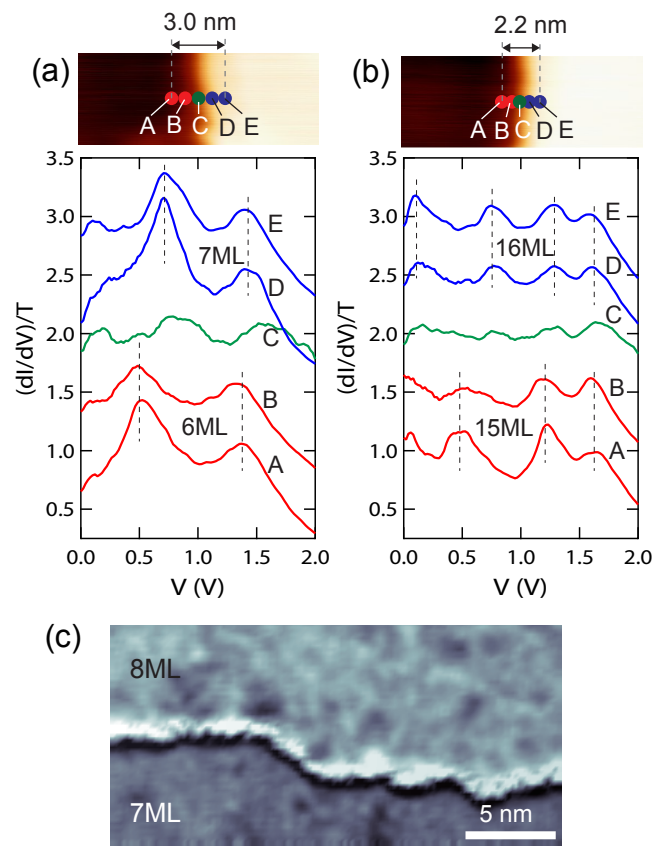

Fig. 4

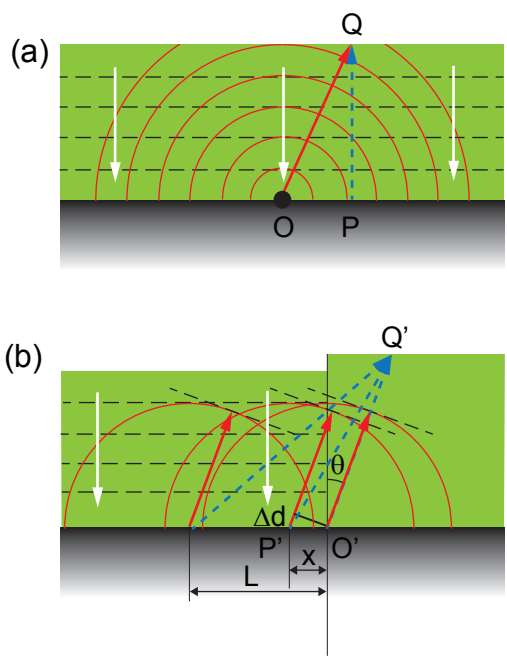

Fig. 5